\documentclass[a4paper,10pt]{article}[2012/07/29]
\usepackage[cp1251]{inputenc}
\usepackage[english]{babel}
\usepackage{amsmath}
\usepackage{amssymb}
\usepackage{graphicx}
\usepackage[pdftex]{color}
\usepackage[sort&compress,square,numbers,comma]{natbib}
\usepackage[pdftex]{hyperref}
\hypersetup{
             final,
             colorlinks=true,
             linkcolor=blue,
             citecolor=red
}

\begin{document}
\twocolumn[\scriptsize{\slshape ISSN 1063-7834, Physics of the Solid State, 2010, Vol. 52, No. 8, pp. 1763–1767. \textcopyright\, Pleiades Publishing,
Inc., 2010.}

\scriptsize{\slshape Original Russian Text \textcopyright\, P.V. Ratnikov, A.P.  Silin, 2010, published in Fizika Tverdogo Tela, 2010,  Vol. 52, No. 8, pp. 1639–1643.}

\vspace{.75cm}

\rule{4.7cm}{.5pt}\hspace{7.6cm}\rule{4.7cm}{.5pt}

\vspace{-.17cm}

\rule{4.7cm}{.5pt}\hspace{7.6cm}\rule{4.7cm}{.5pt}

\begin{center}

\vspace{-1cm}
\large{\bf LOW-DIMENSIONAL SYSTEMS}

\large{\bf AND SURFACE PHYSICS}

\end{center}

\begin{center}
{\fontfamily{ptm}\fontsize{15.8pt}{20pt}\selectfont{\bf Boundary States in Graphene Heterojunctions}}

\normalsize

\vspace{0.3cm}

\large{\bf P.\,V. Ratnikov$^*$ and A.\,P. Silin}

\vspace{0.1cm}

\normalsize

\textit{Lebedev Physical Institute, Russian Academy of Sciences,}

\textit{Leninski\u{\i} pr. 53, Moscow, 119991 Russia}

\textit{$^*$e-mail: \url{ratnikov@lpi.ru}}

Received December 8, 2009; in final form, January 20, 2010
\end{center}

\vspace{0.1cm}
\begin{list}{}
{\rightmargin=1cm \leftmargin=1cm}
\item
\small{{\bf Abstract}---A new type of states in graphene-based planar heterojunctions has been studied in the envelope wave function approximation. The condition for the formation of these states is the intersection between the dispersion curves of graphene and its gap modification. This type of states can also occur in smooth graphene-based heterojunctions.}

\vspace{0.05cm}

\small{\bf DOI}: 10.1134/S1063783410080305

\end{list}\vspace{0.5cm}]

Graphene is a promising material for future carbon nanoelectronics. Owing to its unique electronic properties, this material has attracted particular attention of both theoreticians and experimentalists. For example, the mobility of charge carries in graphene can be as high as $2\cdot10^5$ cm$^2$/V$\times$s and the transport in submicron samples can be ballistic \cite{Du, Morozov}.

We consider a planar heterojunction composed of graphene and a gap modification of graphene. When we say a gap modification of graphene we imply a graphene with an energy gap in the Dirac spectrum of charge carriers. There are several gap modifications of graphene.

First, the energy gap can open because graphene sheets are located not on SiO$_2$ substrate but on some other material, for example, hexagonal boron nitride (h-BN), when two triangular sublattices of graphene become nonequivalent and a gap modification of graphene is formed with an energy gap of 53 meV \cite{Giovannetti}. Second, the energy gap opens in the epitaxially grown graphene on the SiC substrate \cite{Mattausch}, which is equal to 0.26 eV according to experimental results obtained by angular-resolved photoemission spectroscopy \cite{Zhou1}. Third, recently another modification of graphene, i.e., graphane, was synthesized by hydrogenation \cite{Elias}, which has a direct energy gap of 5.4 eV at the $\Gamma$ point according to the calculations \cite{Lebegue}. In the first two cases, a graphene film deposited on inhomogeneous SiO$_2$--h-BN or SiO$_2$--SiC substrates can be used (\hyperlink{Fig1}{Fig. 1a} shows the case with h-BN). In the third case, an inhomogeneously hydrogenated graphene is used (a part of the graphene sample is left without hydrogenation, \hyperlink{Fig1}{Fig. 1b}).

We assume that the energy gap in the gap modifications of graphene opens at $K$ and $K^\prime$ points of the first Brillouin zone, which correspond to the Dirac points of gapless graphene.

Let us assume that the $x$ axis is directed along the plane of the heterojunction perpendicular to the boundary between graphene and its gap modification and the $y$ axis is directed along the boundary. The $z$ axis is directed perpendicular to the plane of the heterojunction. The half-plane  $x$ $<$ 0 is occupied by the gap-less graphene and the half-plane $x$ $>$ 0 belongs to the gap modification of graphene.

\begin{figure}[!b]
\hypertarget{Fig1}{}
\includegraphics[width=0.5\textwidth]{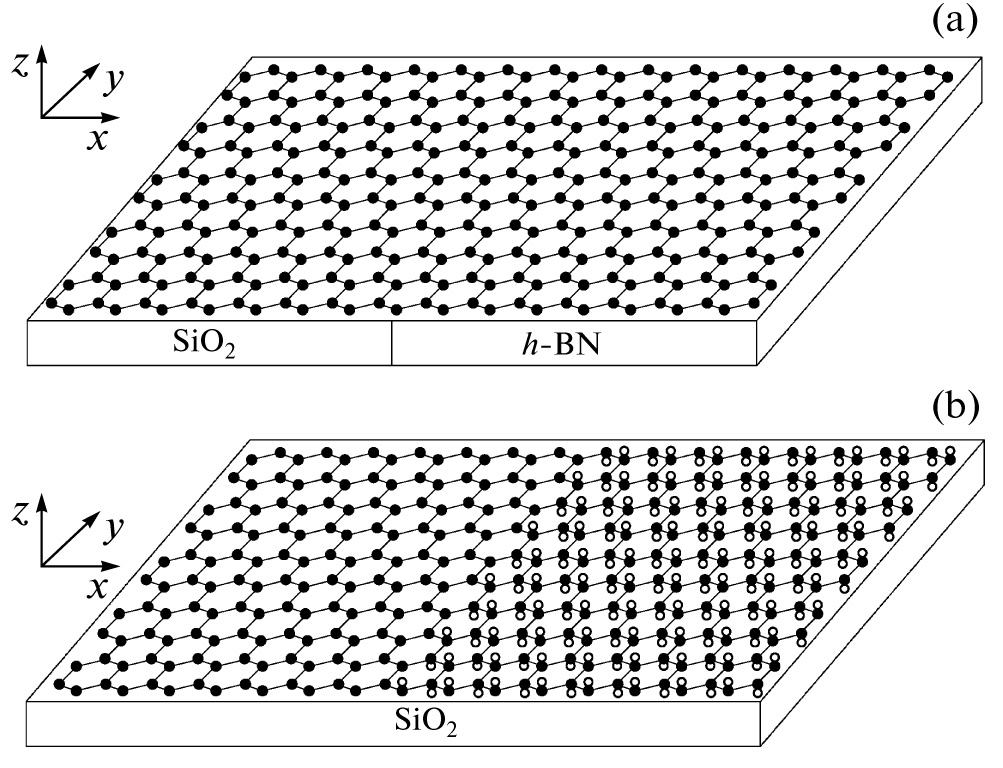}
{\bf Fig. 1.} Two variants of the system in question: {\bf(a)}~graphene layer on the substrate composed of h-BN and SiO$_2$ and {\bf(b)} nonuniformly hydrogenated graphene on the SiO$_2$ substrate. Open circles are  hydrogen atoms, which are located so that they are bound to carbon atoms of one sublattice on one side of graphene sheet and to carbon atoms of the other sublattice on the other side.
\end{figure}

The equation for the envelope wave function that describes charge carriers in the planar graphene-based heterojunction is written in the form \cite{Suzuura, Ando}
\begin{equation}\label{1}
\begin{split}
&\left[v_{Fj}\left(\tau_0\otimes\sigma_x\widehat{p}_x+\tau_z\otimes\sigma_y\widehat{p}_y\right)+\tau_0\otimes\sigma_z\Delta_j\right.\\ &+\left.\tau_0\otimes\sigma_0\left(V_j-E\right)\right]\Psi(x,\,y)=0,
\end{split}
\end{equation}
Here, the parameters with $j$ = 1 are related to the gapless graphene and the parameters with $j$  = 2 are related to the gap modification of the graphene: $v_{F1}$ and $v_{F2}$ are the Fermi velocities (in the general case, $v_{F2}$ $\neq$ $v_{F1}$, and $v_{F1}$ $\approx$ 10$^8$ cm/s); $\Delta_1=0$ and $\Delta_2\neq0$ are the half-widths of the energy gaps; $V_1$ and $V_2$ are the work functions ($V_2$ determines position of the middle of the energy gap for the gap modification of the graphene with respect to the Dirac points of the gapless graphene, and $V_1$ = 0 is chosen for the origin, see \hyperlink{Fig2}{Fig. 2}). The Pauli matrices $\sigma_x,\,\sigma_y,\,\sigma_z$ and $2\times2$ unit matrix $\sigma_0$ operate in the sublattice space ($A$ and $B$ sublattices of graphene hexagonal lattice). The Pauli matrices $\tau_x,\,\tau_y,\,\tau_z$ and $2\times2$ unit
matrix $\tau_0$ operate in the valley space ($K$ and $K^\prime$ points of the first Brillouin zone). Sign $\otimes$ denotes the Kronecker matrix product. The operators $\widehat{p}_x=-i\frac{\partial}{\partial x}$ and $\widehat{p}_y=-i\frac{\partial}{\partial y}$ are the momentum operators ($\hbar=1$).

\begin{figure}[!t]
\hypertarget{Fig2}{}
\includegraphics[width=0.5\textwidth]{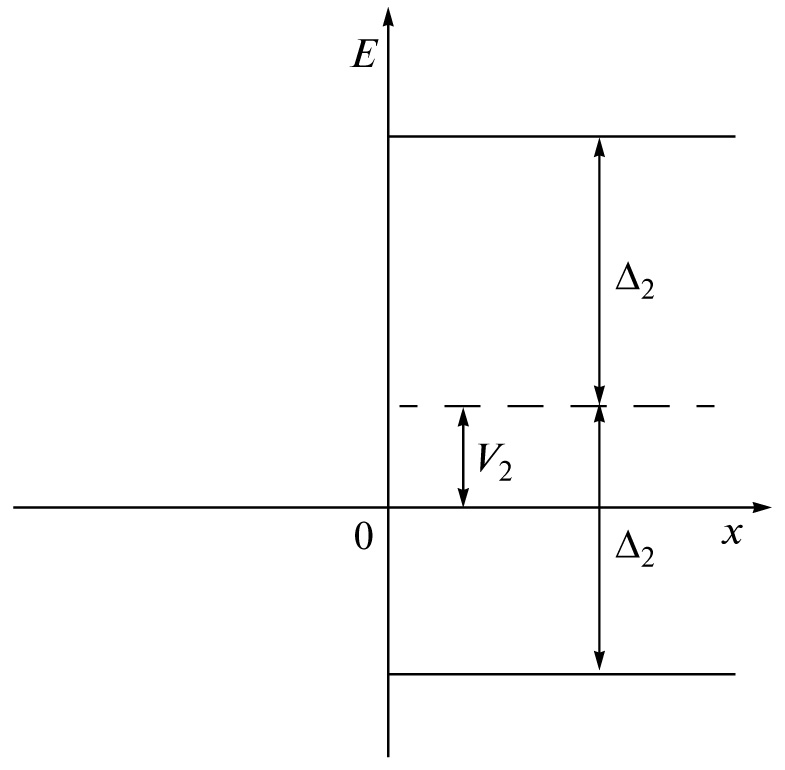}
{\bf Fig. 2.} Graphene heterojunction under consideration.
\end{figure}

In order to avoid spontaneous generation of electron–hole pairs, we assume that the heterojunction in question is a junction of the first kind, i.e., the Dirac points of gapless graphene are located inside the energy gap of its gap modification. This limits value of the work function $|V_2|<\Delta_2$.

Motion of charge carriers along y axis is free:
\begin{equation}\label{2}
\Psi(x,y)=\Psi(x)e^{ik_yy}.
\end{equation}
The wave function $\Psi(x)$ is a bispinor
\begin{equation*}
\Psi(x)=\begin{pmatrix}\Psi_K(x)\\ \Psi_{K^\prime}(x)\end{pmatrix},
\end{equation*}
where $\Psi_K(x)$ and $\Psi_{K^\prime}(x)$ spinors describe charge carriers in the $K$ and $K^\prime$ valleys, respectively:
\begin{equation*}
\Psi_K(x)=\begin{pmatrix}\psi_{KA}(x)\\ \psi_{KB}(x)\end{pmatrix},\hspace{0.2cm}\Psi_{K^\prime}(x)=\begin{pmatrix}\psi_{K^\prime A}(x)\\ \psi_{K^\prime B}(x)\end{pmatrix}.
\end{equation*}

Let us consider the parity operator
\begin{equation}\label{3}
\widehat{P}=\tau_z\otimes\sigma_0,
\end{equation}
which is a product of the inversion operator $i\gamma^\prime_4=i\tau_z\otimes\sigma_z$ and the operator of rotation by the angle $\pi$ about the $z$ axis $\widehat{\Lambda}_z=-i\tau_0\otimes\sigma_z$. Evidently, the operator \eqref{3} commutes with the Hamiltonian in Eq. \eqref{1}.

Equation \eqref{1} is solved within the class of wave eigenfunctions $\Psi_\lambda(x)$ of the parity operator \eqref{3}
\begin{equation}\label{4}
\begin{split}
\widehat{P}\Psi_\lambda(x)=\lambda\Psi_\lambda(x),&\hspace{0.07cm}\lambda=\pm1,\\
\Psi_{+1}(x)=\begin{pmatrix}\Psi_{+1,K}(x)\\
0\end{pmatrix},&\hspace{0.07cm} \Psi_{-1}(x)=\begin{pmatrix}0\\ \Psi_{-1,K}(x)\end{pmatrix}.
\end{split}
\end{equation}
Equation \eqref{1} can be easily represented as two $2\times2$ matrix equations
\begin{equation}\label{5}
\begin{split}
&\left(-iv_{Fj}\sigma_x\frac{d}{dx}+v_{Fj}k_y\sigma_y+\lambda\Delta_j\sigma_z+V_j\right)\Psi_{\lambda K}(x)\\ &=E_\lambda\Psi_{\lambda K}(x),
\end{split}
\end{equation}
\begin{equation}\label{6}
\begin{split}
&\left(-iv_{Fj}\sigma_x\frac{d}{dx}-v_{Fj}k_y\sigma_y-\lambda\Delta_j\sigma_z+V_j\right)\Psi_{\lambda K^\prime}(x)\\ &=E_\lambda\Psi_{\lambda K^\prime}(x).
\end{split}
\end{equation}
In this case, we have $\lambda$ = +1 in Eq. \eqref{5} and $\lambda$ = –1 in Eq. \eqref{6}.

Clearly, for $\Delta_j$ = 0 and $V_j$  = 0, we return to the spinor wave functions that describe the chiral states either ar the $K$ point or at the $K^\prime$ point. In this case, it is possible to introduce helicity operator $\widehat{h}$ = ${\boldsymbol\sigma}\cdot{\bf p}/(2|{\bf p}|)$. Its eigenvalue (helicity) determines the attribution of charge carriers to one of two valleys \cite{Neto}. However, for $\Delta$ $\neq$ 0, the chiral symmetry is broken, and, therefore, instead of the helicity the quantum number $\lambda$ (parity) is introduced, which, according to Eq. \eqref{4}, determines attribution of charge carriers to one of the two valleys.

We use the following condition of matching the envelope wave functions \cite{Kolesnikov1, Silin}
\begin{equation}\label{7}
\sqrt{v^{(-)}_F}\Psi^{(-)}_\lambda=\sqrt{v^{(+)}_F}\Psi^{(+)}_\lambda,
\end{equation}
where the signs ``$-$'' and ``+'' indicate the quantities related to the material on the left-hand and right-hand sides of the interface, respectively.

The solution to Eq. \eqref{5} for boundary states has the form
\begin{equation}\label{8}
\Psi_{\lambda K}(x)=\begin{cases}C{1\choose a}\exp({\kappa_1x}),&x<0,\\
C{b\choose qb}\exp({-\kappa_2x}),&x>0,\end{cases}
\end{equation}
where
\begin{equation*}
a=i\frac{v_{F1}(k_y-\kappa_1)}{E_\lambda},\hspace{0.25cm}q=i\frac{v_{F2}(k_y+\kappa_2)}{E_\lambda-V_2+\lambda\Delta_2},
\end{equation*}
$C$ is the normalization factor, $b=\sqrt{\frac{v_{F1}}{v_{F2}}}$ is the constant obtained when matching solutions for $x$ $<$ 0 and $x$~$>$~0 at the line $x$ = 0 under condition \eqref{7},
\begin{equation}\label{9}
E_\lambda=\pm v_{F1}\sqrt{k^2_y-\kappa^2_1},
\end{equation}
from which it follows that the necessary condition for the existence of the boundary states is given by inequality
\begin{equation}\label{10}
\kappa_1<|k_y|.
\end{equation}
Equation \eqref{9} can be rewritten as
\begin{equation*}
\kappa_1=\sqrt{k^2_y-E^2_\lambda/v^2_{F1}},
\end{equation*}
Therefore, the following inequality should also be
valid
\begin{equation}\label{11}
|E_\lambda|<v_{F1}|k_y|.
\end{equation}
Expression for $\kappa_2$ is represented in the form
\begin{equation*}
\kappa_2=\frac{1}{v_{F2}}\sqrt{\Delta^2_2-(E_\lambda-V_2)^2+v^2_{F2}k^2_y}.
\end{equation*}
Moreover, the matching leads to the inequality
\begin{equation}\label{12}
\frac{v_{F1}(k_y-\kappa_1)}{E_\lambda}=\frac{v_{F2}(k_y+\kappa_2)}{E_\lambda-V_2+\lambda\Delta_2}.
\end{equation}
The solution to Eq. \eqref{6} is produced from Eq. \eqref{8} by the following substitutions in factors $a$ and $q$: $k_y\rightarrow-k_y$ and $\lambda\rightarrow-\lambda$.

Let us discuss separately the case of zero mode $E_\lambda=0$. Components of the envelope wave function in $x$ $<$ 0 region (gapless graphene) $\Psi_{\lambda K}={a_1 \choose a_2}\exp(\kappa_1x)$ satisfy equations:
\begin{equation*}
\begin{split}
(\kappa_1-k_y)a_1&=0,\\
(\kappa_1+k_y)a_2&=0,
\end{split}
\end{equation*}
i.e., either $\kappa_1$~=~$k_y$ ($k_y>0$) and $a_2=0$, or $\kappa_1$~=~$-k_y$
($k_y<0$) and $a_1=0$. Then it follows from the matching condition \eqref{7} that both components of the envelope wave function are zero in $x$ $>$ 0 region ($b$ = 0); therefore, we have $a_1=0$ and $a_2=0$, i.e., $\Psi_{\lambda K}(x)$ $\equiv$ 0. Thus, there is no zero mode for the boundary states in question.

The following equations are easily obtained from Eq. \eqref{12}:
\begin{equation}\label{13}
\kappa_1\kappa_2=\frac{E_\lambda(E_\lambda-V_2)}{v_{F1}v_{F2}}-k^2_y,
\end{equation}
\begin{equation}\label{14}
\lambda\Delta_2E_\lambda=v_{F1}v_{F2}k_y(\kappa_1+\kappa_2).
\end{equation}
The two latter equations are valid for either value of $\lambda$ (for both valleys), because they are invariant in respect to simultaneous substitutions $k_y$ $\rightarrow$ $-k_y$ and $\lambda$ $\rightarrow$ $-\lambda$.

Since $\kappa_1>0$ and $\kappa_2>0$, right-hand side of Eq.~\eqref{13} should be positive. Let us denote by $\varepsilon_0(k_y)$ such value of $E_\lambda$ that the right-hand side of Eq. \eqref{13} turns zero,
\begin{equation}\label{15}
\varepsilon_0(k_y)=\frac{V_2}{2}\pm\sqrt{\frac{V^2_2}{4}+v_{F1}v_{F2}k^2_y},
\end{equation}
where ``+'' corresponds to electrons and ``$–$'' to holes.
Then, the condition $\kappa_1\kappa_2>0$ is equivalent to the inequality
\begin{equation}\label{16}
|E_\lambda|>|\varepsilon_0(k_y)|.
\end{equation}

It follows from Eq. \eqref{14} that inequality $\lambda k_y$~$>$~0 holds for electron boundary states ($E_\lambda>0$), and $\lambda k_y$~$<$~0 holds for hole boundary states ($E_\lambda<0$). The boundary states are not degenerate in parity. That means that there is no Kramers degeneracy of energy spectrum for them. This is also true for boundary states in a planar quantum well based on graphene nanoribbon \cite{Ratnikov} and for boundary states localized on zigzag edges of gapless graphene \cite{Tkachov}. Since parity determines charge carrier attribution to one of two valleys, the property mentioned above means also that there is a ``valley polarization'' of boundary states: electrons that move along the heterojunction boundary with $k_y>0$ are located near $K$ point and electrons with $k_y<0$ are near $K^\prime$ point and vise versa in case of holes. Because of that, current that flows along the heterojunction boundary would be ``valley-polarized''.

By squaring Eq. \eqref{14} we get a quadratic equation, solution of which produces dependence of energy on $k_y$:
\begin{equation}\label{17}
E_\lambda(k_y)=\frac{v_{F1}v_{F-}k^2_yV_2+\lambda v_{F1}k_y\Delta_2\sqrt{\Delta^2_2+v^2_{F-}k^2_y-V^2_2}}{\Delta^2_2+v^2_{F-}k^2_y},
\end{equation}
where $v_{F-}=v_{F1}-v_{F2}$. Equation \eqref{17} takes into account that sign of $\lambda k_y$ determines type of charge carriers in the boundary states.

It is easy to verify that inequality \eqref{11} is always true if the energy is given by Eq. \eqref{17}. Therefore, inequality \eqref{10} also holds.

Now, it is simple to analyze inequality \eqref{16}. Let us introduce the following notation:
\begin{equation*}
k_{y1}=\frac{|V_2|}{|v_{F-}|},
\end{equation*}
\begin{equation*}
k_{y2,\,3}=\sqrt{\frac{v_{F2}V^2_2+2v_{F-}\Delta^2_2\mp|V_2|\sqrt{v^2_{F2}V^2_2+4v_{F1}v_{F-}\Delta^2_2}} {2v_{F2}v^2_{F-}}}.
\end{equation*}
Under the condition
\begin{equation*}
v_{F1}\,<\,v_{F2}\,<\,2v_{F1},\hspace{0.25cm} \frac{2}{v_{F2}}\sqrt{v_{F1}|v_{F-}|}\Delta_2\,<\,|V_2|\,<\,\Delta_2,
\end{equation*}
the boundary states exist in the ranges\footnote{Here and below, we exclude the point $k_y=0$, because it corresponds to $E_\lambda=0$.}
\begin{equation*}
0<|k_y|<k_{y2},\hspace{0.25cm} k_{y3}<|k_y|<k_{y1}
\end{equation*}
either for electrons, if $V_2<0$, or for holes, if $V_2>0$.

Under condition
\begin{equation*}
v_{F1}\,<\,v_{F2}\,<\,2v_{F1},\hspace{0.25cm} |V_2|\,<\,\frac{2}{v_{F2}}\sqrt{v_{F1}|v_{F-}|}\Delta_2
\end{equation*}
the boundary states exist in the range
\begin{equation*}
0<|k_y|<k_{y1}
\end{equation*}
either for electrons, if $V_2<0$, or for holes, if $V_2>0$.

Under the condition
\begin{equation*}
v_{F1}\,>\,v_{F2},\hspace{0.25cm} 0\,<\,V_2\,<\,\Delta_2
\end{equation*}
the electron boundary states exist in the range
\begin{equation*}
k_{y3}<|k_y|<k_{y1},
\end{equation*}
and the hole boundary states exist in the range
\begin{equation*}
0<|k_y|<k_{y2}.
\end{equation*}
Under condition
\begin{equation*}
v_{F1}\,>\,v_{F2},\hspace{0.25cm} -\Delta_2\,<\,V_2\,<\,0
\end{equation*}
the electron boundary states exist in the range
\begin{equation*}
0<|k_y|<k_{y2},
\end{equation*}
and the hole boundary states exist in the range
\begin{equation*}
k_{y3}<|k_y|<k_{y1}.
\end{equation*}

Let us consider three special cases.

(1) Under condition $V_2$ = 0 and $v_{F-}$ $\neq$ 0, the boundary states exist for both electrons and holes in the following range if $v_{F1}$ $>$ $v_{F2}$
\begin{equation*}
0<|k_y|<\frac{\Delta_2}{\sqrt{v_{F1}v_{F-}}}.
\end{equation*}

(2) Under condition $v_{F1}$ = $v_{F2}$, 0 $<$ $|V_2|$ $<$ $\Delta_2$ the
boundary states exist in the range
\begin{equation*}
0<|k_y|<\frac{\Delta_2\sqrt{\Delta^2_2-V^2_2}}{v_{F2}|V_2|}
\end{equation*}
either for electrons, if $V_2<0$, or for holes, if $V_2>0$.

(3) Under condition $v_{F1}$ = $v_{F2}$, $V_2=0$, the boundary states are absent both for electrons and holes, because $|E_\lambda(k_y)|=|\varepsilon_0(k_y)|$, which is in contradiction with inequality \eqref{16}.

\hyperlink{Fig3}{Fig. 3} shows dispersion curves $E^{e,h}_\lambda(k_y)$ and $\varepsilon^{e,h}_0(k_y)$ for the electron and hole boundary states for three values of $V_2$ in the model of graphene-based heterojunction with $\Delta_2$ = 260 meV and $v_{F2}$ = 1.2 $\times$ $10^8$ cm/s for gap modification of graphene.

Our results remain in essence the same if instead of a sharp heterojunction we consider a smooth heterojunction. Indeed, let $v_F(x)$ and $\Delta(x)$ vary smoothly from their values for gapless graphene to their values in gap modification of graphene over a strip with the width $d$ $\lesssim$ $\kappa^{-1}_{1,\,2}$. Then change in energy of the boundary states is $|\delta E_\lambda(k_y)|$ $\lesssim$ 1 meV. Such insignificant variation in energy of the boundary states produces no noticeable qualitative changes. A similar result has been obtained for boundary states in heterojunctions of narrow-gap semiconductors with intercrossing dispersion curves in \cite{Kolesnikov2}.

To conclude, we would like to point out that the new type of boundary states in graphene heterojunctions can be studied in experiment by tunnel spectroscopy of angular-resolved photoemission spectroscopy similar to how it have been done for boundary states in gapless graphene \cite{Kobayashi, Niimi, Zhou2}.

\begin{figure}[!t]
\hypertarget{Fig3}{}
\includegraphics[width=0.5\textwidth]{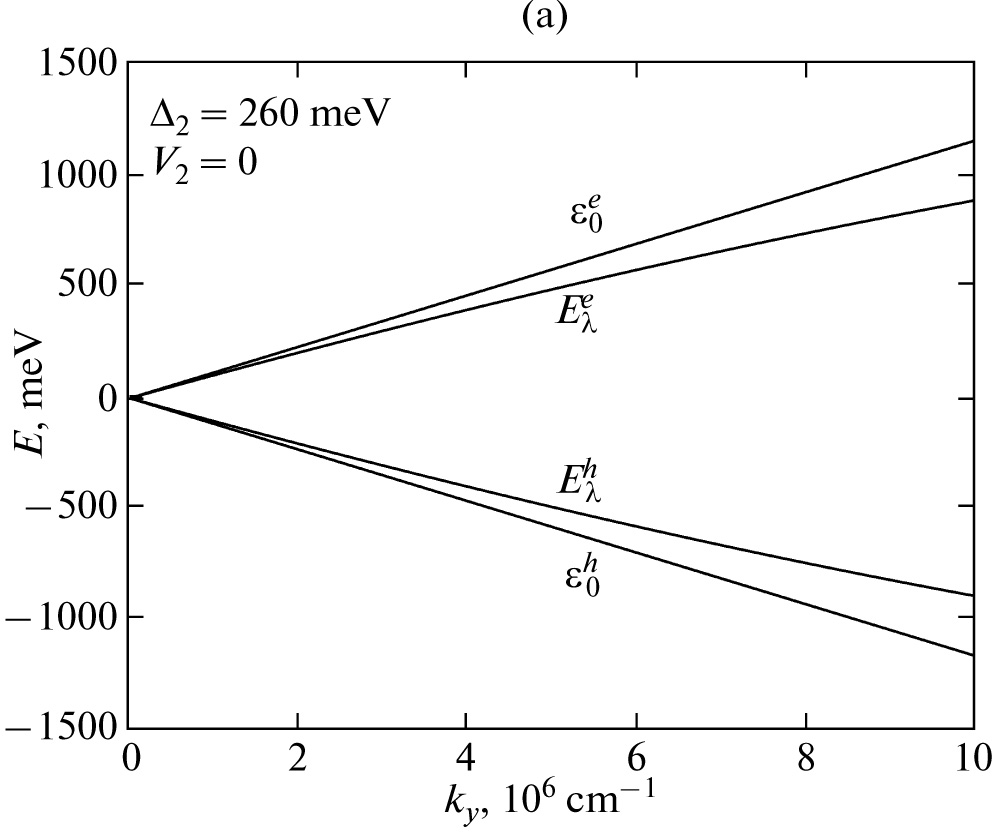}\\
\includegraphics[width=0.5\textwidth]{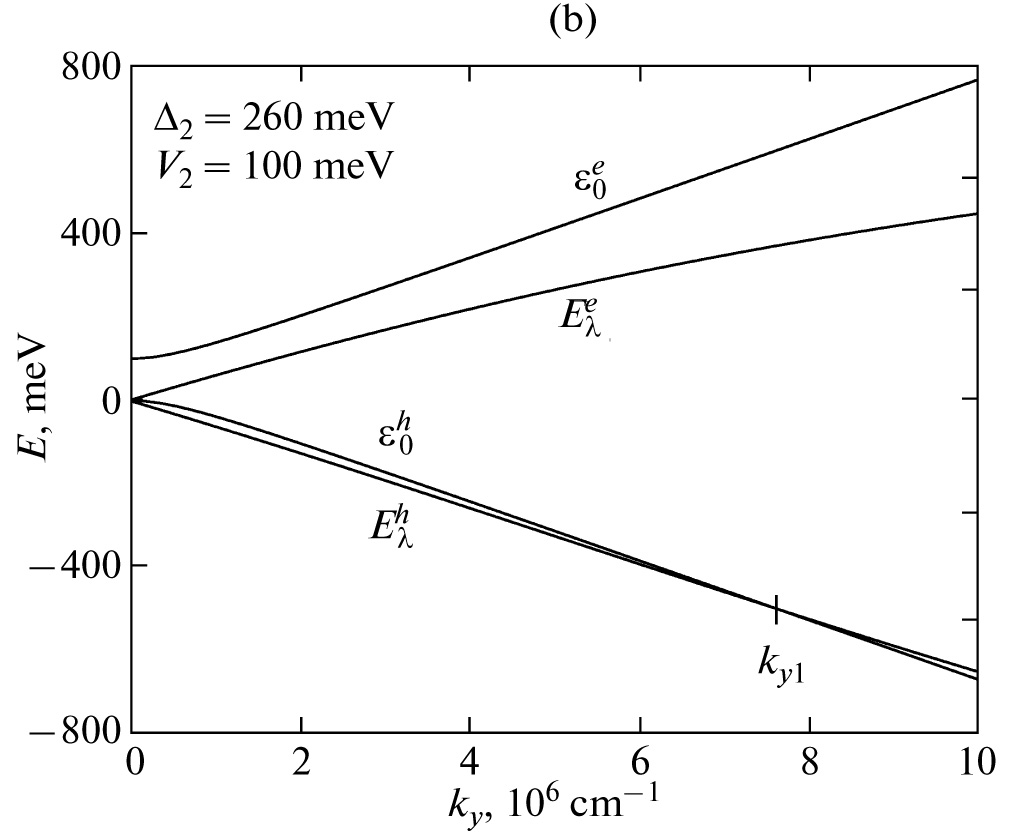}\\
\includegraphics[width=0.5\textwidth]{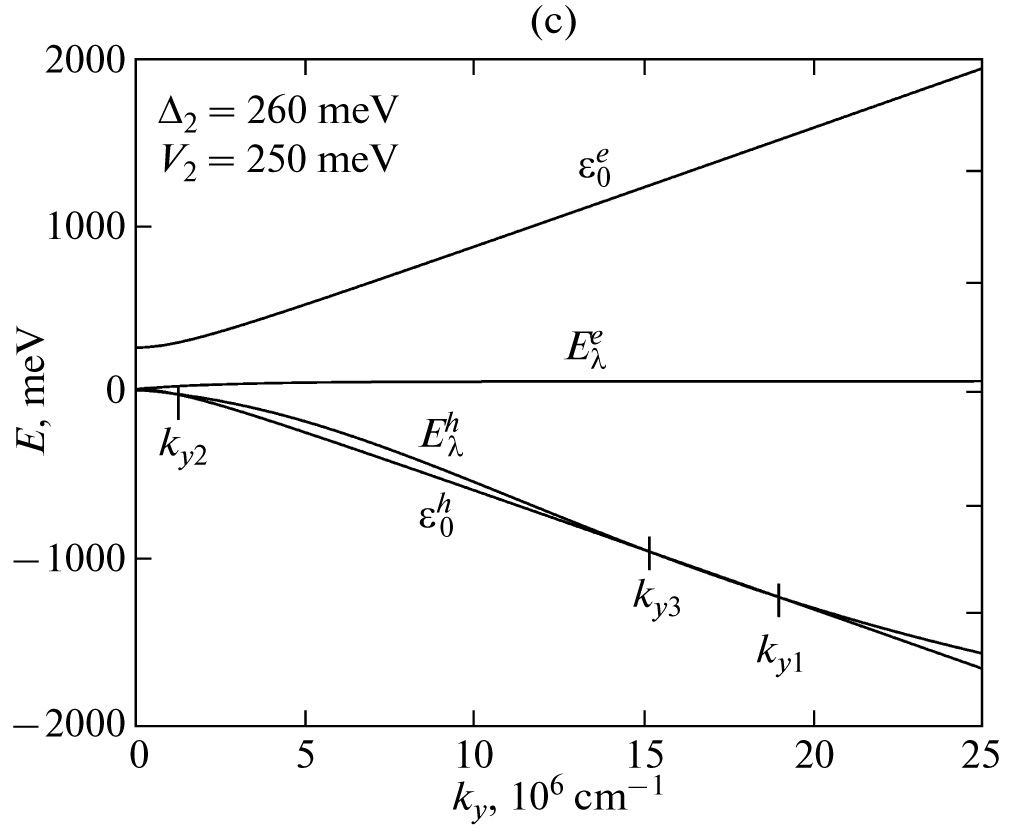}\\
\vspace{0.25cm}
{\bf Fig. 3.} Dispersion curves $E^{e,h}_\lambda(k_y)$ and $\varepsilon^{e,h}_0(k_y)$: {\bf(a)}~there are no  boundary states for electrons and holes at $V_2=0$, {\bf(b)} there are only hole boundary states in the range $0<|k_y|<k_{y1}$ at $V_2$ = 100 meV, and {\bf(c)} there are only  hole boundary states in the ranges $0<|k_y|<k_{y2}$ and $k_{y3}<|k_y|<k_{y1}$ at $V_2$ = 250 meV.
\end{figure}

\begin{center}
ACKNOWLEDGMENTS
\end{center}

This study was supported in part by the Dynasty Foundation, the Center for Science and Education of the Lebedev Physical Institute of the Russian Academy of Sciences, and the Presidium of the Russian Academy of Sciences (program for support of young scientists).

\end{document}